\documentclass[aps, pra,twocolumn, showpacs,
superscriptaddress,10pt,floatfix,longbibliography]{revtex4-2}

\usepackage{graphicx}
\usepackage{grffile}
\usepackage{amsthm}
\usepackage{epsfig}
\usepackage{amsmath,amssymb,amsfonts,mathtools,bbm,MnSymbol,mathrsfs,xfrac}
\usepackage{longtable}
\usepackage{tabularx}
\usepackage{colortbl}
\usepackage[table]{xcolor}
\usepackage[breaklinks=true,colorlinks=true,linkcolor=blue,urlcolor=blue,citecolor=blue]{hyperref}
\usepackage{comment}
\usepackage{color}
\usepackage[makeroom]{cancel}
\usepackage{tikz}
\usetikzlibrary{positioning}
\usetikzlibrary{arrows.meta}

\usepackage{soul}
\pdfoutput=1

\newcommand{\M}{\mathcal{M}}

\renewcommand{\[}{\begin{equation}}
\renewcommand{\]}{\end{equation}}

\newcommand{\ket}[1]{|#1\rangle}

\newcommand{\braket}[2]{\langle#1|#2\rangle}
\newcommand{\pro}[2]{|#1\rangle\langle#2|}
\newcommand{\mean}[1]{\langle#1\rangle}
\newcommand{\abs}[1]{|#1|}

\newcommand{\tr}{\mathrm{tr}}

\newcommand{\R}{{\rho}}

\newcommand{\I}{{I}}

\newcommand{\C}{{\mathcal{C}}}

\newcommand{\Svn}{{S}}

\renewcommand{\P}{{P}}

\renewcommand{\H}{{H}}

\newcommand{\cg}{{\mathrm{cg}}}

\newcommand{\sect}[1]{\medskip\noindent \textbf{#1. }}

\definecolor{mygray}{gray}{0.6}

\theoremstyle{definition}

\definecolor{dfcol}{cmyk}{1, 0.2108, 0.13, 0.3}
\newcommand{\df}[1]{\ifthenelse{\boolean{}}{\textcolor{dfcol}{[{\bf DF}: #1]}}{}}

\begin{document}


\title{Ergotropic interpretation of entanglement entropy}

\author{Dominik \v{S}afr\'{a}nek}
\email{dsafranekibs@gmail.com}
\affiliation{Center for Theoretical Physics of Complex Systems, Institute for Basic Science (IBS), Daejeon - 34126, Korea}

\date{\today}

\begin{abstract}
Entanglement entropy is one of the most prominent measures in quantum physics. We show that it has an interesting ergotropic interpretation in terms of unitarily extracted work. It determines how much energy one can extract from a source of pure unknown states by applying unitary operations when only local measurements can be performed to characterize this source. Additionally, entanglement entropy sets a limit on the minimal temperature to which these partially characterized states can be cooled down, by using only unitary operations.
\end{abstract}

\maketitle

Entanglement entropy is one of the most used entropy measures in quantum physics and certainly the most used measure of quantum entanglement. 
It is defined for a pure bipartite state $\ket{\psi_{AB}}$ as the von Neumann entropy of the reduced state of either of the subsystems~\cite{horodecki2009quantum,nielsen2010quantum,eisert2010colloquium,nishioka2018entanglement,headrick2019lectures},
\[
S_{\mathrm{ent}}=\Svn(\R_A)=\Svn(\R_B),
\]
where $\R_A=\tr_B[\R]$, $\R_B=\tr_A[\R]$, and $\R=\pro{\psi_{AB}}{\psi_{AB}}$. It describes the loss of information due to having access to only one of the two subsystems. Because the global state is pure, this loss of information must be caused solely by the presence of entanglement. Thus, entanglement entropy measures how much entanglement the subsystem $A$ shares with subsystem $B$.

Its usefulness as an entanglement measure has been widely appreciated~\cite{plenio2007introduction} since entanglement is the most significant feature that distinguishes quantum from classical physics and the primary resource of quantum devices such as quantum computers~\cite{jozsa1997ent,cao2018quantum,aaronson2008limits}, quantum sensors~\cite{giovannetti2004quantum,giovannetti2011advances,zhang2015entanglement}, and cryptographical devices~\cite{pirandola2020advances,elliott2005current,chen2021integrated}. 

Its prominence, however stems from its applications in many-body physics~\cite{eisert2010colloquium,abanin2019colloquium}, quantum field theory~\cite{Pasquale2004entanglement,nishioka2018entanglement}, and cosmology~\cite{das2008black,Borsten_2012}, fields which are somewhat surprisingly connected through this quantity. In many-body physics, its power lies in analyzing the amount of entanglement of typical eigenstates of the Hamiltonian and its scaling with the system size and the dimension. It has been found that entanglement typically scales either with the dimension of the Hilbert space (known as volume law) or with the size of the boundary that separates the two subsystems (known as area law),  informing on its thermalization properties~\cite{eisert2010colloquium,abanin2019colloquium}. The volume law is usually followed by random or high-energy eigenstates~\cite{page1993,foong1994proof,sen1996average,alba2009entanglement,ares2014excited,storms2014entanglement,keating2015spectra,miao2021eigenstate}, in which entanglement entropy equals the thermodynamic entropy of the smaller subsystem as a consequence of the eigenstate thermalization hypothesis~\cite{deutsch1991ETH,srednicki1994ETH,Deutsch_2018}. There are some deviations in various circumstances~\cite{vidmar2017entanglement,vidmar2017entanglement,faiez2020typical,faiez2020how}. Ground states typically follow area law~\cite{bombelli1986quantum,srednicki1993entropy}, which was proved for a ground state of any gapped Hamiltonian~\cite{Hastings_2007}. Low entanglement can also indicate the presence of many-body quantum scars~\cite{turner2017quantum,chatto2020quantum,desaules2022extensive}, exceptional energy eigenstates that have a surprisingly low amount of entanglement compared to other energy eigenstates. This leads to an interesting phenomenon: for some specific initial states, the quantum scar system, which otherwise thermalizes well, suddenly shows recurrences.

Having low entanglement in the ground state means that such states can be efficiently simulated by the Density matrix renormalization group (DMRG)~\cite{schollwock2008density,chan2011density} and other methods (PEPS~\cite{verstraete2006criticality}, MERA~\cite{cincio2008multiscale}) developed in past years. This is important because ground states can encode solutions to many-interesting problems, such as the Riemann hypothesis~\cite{bender2017hamiltonian}. The ability of ground states to encode such solutions has been the primary motivation behind quantum annealing and adiabatic quantum computing~\cite{benedetti2016estimation,pino2020mediator,hauke2020perspectives,imoto2021improving,mohseni2019error}. Additionally, these methods can be applied to find the low-energy properties of materials so far applied to spin chains~\cite{alcaraz2011entanglement,couvreur2017entanglement,bayat2022entanglement}.

Entanglement entropy has also been studied as an indicator of quantum phase transitions~\cite{sachdev1999quantum,vidal2003entanglement,vojta2003quantum,hur2008entanglement,ban2002testing,iadecola2022dynamical,jafa2022entanglement,kawabata2023entanglement,pavigli2023multipartite} and as a measure of quantum chaos~\cite{wang2004entanglement,asplund2016entanglement,vidmar2017entanglement,bianchi2018linear,piga2019quantum,lerose2020bridging,vidmar2021entanglement}.

In cosmology, the main interest in entanglement entropy lies in the explanation of the black hole entropy~\cite{bekenstein1973black,bekenstein2004black,hawking1982black,solodukhin2011ent,hartman2013time,nam2022ent,rusalev2022ent,matsuo2022ent,luongo2023ent} and the information paradox~\cite{penington2020entanglement,RAJU20221,Raju_2022,doren2022resolving}. Computing entanglement entropy for quantum fields and finding its expansion coefficients became one of the main tasks. This led to the invention of the replica trick for its analytic computation in 1+1 dimension~\cite{holzey1994geometric}, and holographic principle and AdS/CFT correspondence that lead that provides in higher dimensions~\cite{ryu2006holographic}. These methods then trickled down to the applications in the many-body physics as mentioned above and to a discovery of a related quantity, holographic entanglement entropy~\cite{ryu2006holographic,Nishioka_2009,nishioka2018entanglement}.

Finally, entanglement entropy has been recently studied and applied in machine learning~\cite{deng2017quantum,you2018machine,lam2021machine}.

In all of these applications, the role of entanglement entropy was to serve as a measure of \emph{information}, or rather lack thereof. The role of entropy in thermodynamics is somewhat different. Low entropy indicates that it will increase in the future, according to the second thermodynamic law. This tendency to increase can be utilized productively by devising schemes that direct some energy from the system to a task at hand. This is the basis of any engine in use. In other words, entropy indicates the system's potential to be utilized to perform work, or rather plainly, it measures our ability to extract energy from it. 

In this paper, we provide a thermodynamic interpretation of entanglement entropy. We show that it is related to work extraction in a particular scenario. Consider that we acquire a source of pure but otherwise unknown quantum states, and we have only local measurements to characterize this source, i.e., we can perform only local measurements on subsystems A and B, given the states it produces. In other words, we are not allowed or able to perform a measurement that projects onto states that are entangled between A and B. Entanglement entropy measures how much energy we can extract from this source, given this limited ability to characterize it.

This result will follow from combining two other results. First, a result that relates the characterization of unknown quantum sources, observational entropy, and the amount of energy that can be unitarily extracted. Second, a relation between observational entropy and entanglement entropy. After we establish this main result which relates entanglement entropy with unitary work extraction, we generalize it to include sources of unknown mixed states and conclude.

\begin{figure*}[t!]
\begin{center}
\includegraphics[width=1\hsize]{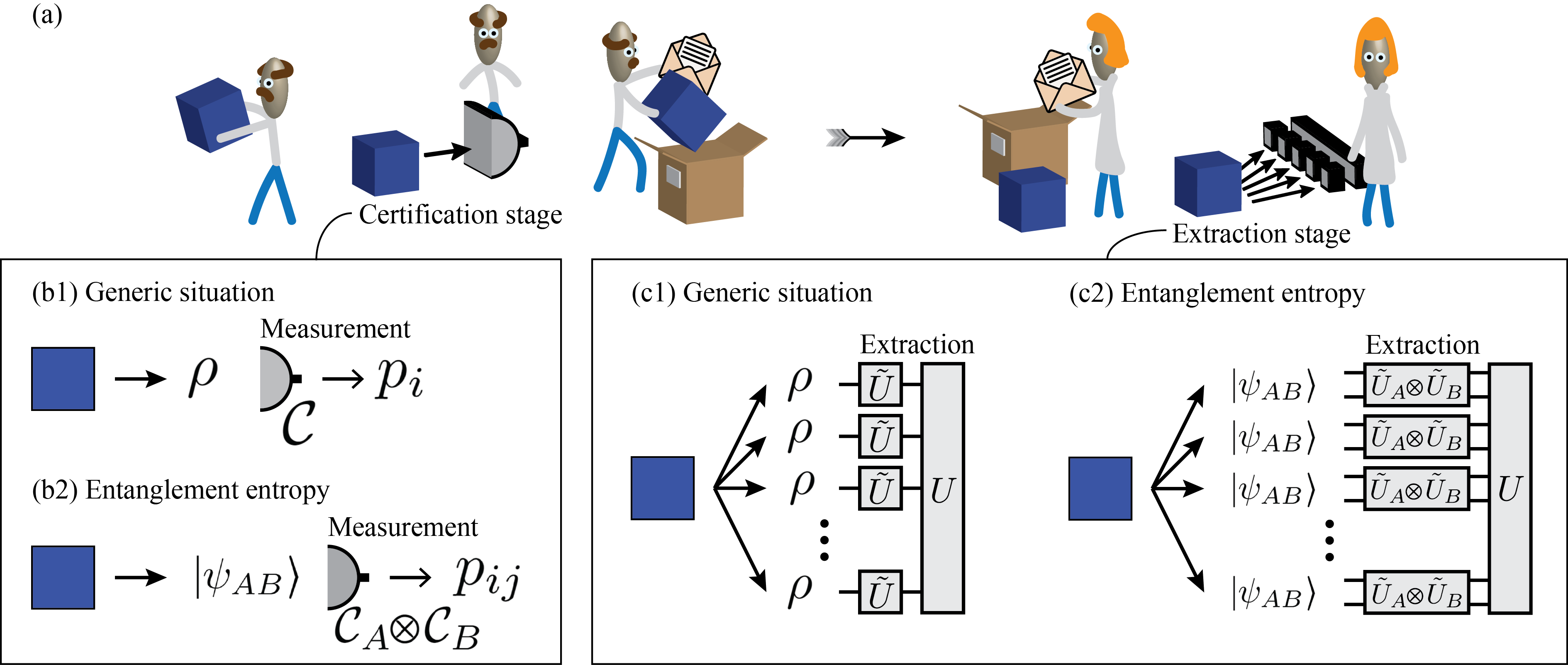}\\
\caption
{{\bf Setup of work extraction from unknown quantum sources.} 
(a) Charlie takes an unknown source (blue cube) and starts making measurements on it. We call this the \emph{certification stage}. He writes a letter that contains information on his measurement basis and the corresponding probabilities of outcomes, which he determined by making many such measurements. He attaches this letter to the source and sends it as a package to Elizabeth. In the \emph{extraction stage}, Elizabeth opens the package and adjusts the random unitaries $\tilde U$ according to the measurement basis she obtained from Charlie. Additionally, she adjusts the final extraction unitary $U$ according to the measurement basis and the probabilities of outcomes she read in the letter. She lets the source produce multiple states before she extracts. She does that because extracting energy from multiple copies simultaneously provides more energy than if she were extracting sequentially from many single copies. If any measurement can be performed during the certification stage (b1), then the maximum extracted work during the corresponding extraction stage (c1) is determined by the von Neumann entropy. If only local measurements can be performed during the certification stage (b2), then the maximum extracted work during the corresponding extraction stage (c2) is given by entanglement entropy if the states produced by the source are pure. If mixed states are produced, with only local measurements available, the maximal extracted work is given by the sum of von Neumann and quantum correlation entropy.
}
\label{Fig:ec}
\end{center}
\end{figure*}

\sect{Review of observational ergotropy} 
The amount of work extractable from a quantum system depends on what is experimentally possible~\cite{Niedenzu2019,kamin2021exergy,lobejko2022work,janovitch2022quantum,morrone2023daemonic,mula2023ergotropy,shaghaghi2022extracting,shaghaghi2022micromasers}. Ergotropy is a measure of work defined as the maximal energy obtained from an isolated quantum system by a unitary extraction~\cite{Allahverdyan2004a,alicki2013entanglement,hovhannisyan2013entanglement,campaioli2018quantum}. In other words, it is the maximal amount by which we can reduce the energy of a quantum system, using only unitary operations. The notion of ergotropy has been discussed in many different contexts and with different assumptions~\cite{Binder2015a,campaioli2017enhancing,alimuddin2019bound,puliyil2022thermodynamic,Brown2016,Francica2020,simonov2022work,biswas2022extraction,culhane2022extractable,salvia2022extracting,koshihara2023quantum,tirone2023quantum}.

Here we will be concerned with a scenario in which the states from which the energy is extracted are unknown but can be characterized by a single coarse-grained measurement~\cite{safranek2023work}. This leads to an alternative notion called \emph{observational ergotropy}.

The main result of~\cite{safranek2023work} can be summarized as follows. Consider two agents, the certifier (Charlie) and the extractor (Elizabeth), which are friendly to each other.

Consider a source of unknown quantum states $\R$ with each state characterized by its Hamiltonian $\H$. If $N$ states are produced, the total Hamiltonian is defined as the sum of the local terms,
\[
\H_N=\H\otimes \I\otimes\cdots+\I\otimes \H\otimes\cdots+\cdots.
\]
See Fig.~\ref{Fig:ec} that illustrates the setup applicable to this paper.

Charlie's goal as a certifier is to measure the unknown states produced by the source in some measurement basis $\{\ket{i}\}$, and determine the probabilities of outcome $p_i$ by doing many of such measurements repeatedly, see Fig.~\ref{Fig:ec} (b1). Then he sends the package of the source together with an accompanying letter to Elizabeth, which states: ``I measured the source in basis $\{\ket{i}\}$ and obtained the probability distribution of outcomes $\{p_i\}$.''

Elizabeth's goal as an extractor is to use the information received from Charlie to extract energy from the source by applying unitary operations. Her strategy is as follows. She will let the source produce $N$ states, then apply a random unitary $\tilde U=\sum_ie^{i \theta_i}\pro{i}{i}$ on each one of them, where angles $\theta_i$ are drawn uniformly from interval $[0,2\pi)$, and finally apply a particular $N$-state global unitary $U$ that depends on the distribution $\{p_i\}$, see Fig.~\ref{Fig:ec} (c1). $U$ is defined as the unitary operator that transforms the product coarse-grained state $\R_{\cg}^{\otimes N}$, $\R_{\cg}=\sum_i p_i\pro{i}{i}$, into its corresponding passive state. Thus, the full extraction unitary that is applied on the product state $\R^{\otimes N}$ is
\[\label{eq:elizabeth_unitary_extraction}
U \tilde U^{\otimes N}.
\]
The work she will extract is measured by the difference between the initial and the final energy of the state,
\[
W_N=\tr[\H_N\R^{\otimes N}]-\tr[\H_NU \tilde U^{\otimes N}\R^{\otimes N}\big(U \tilde U^{\otimes N}\big)^\dag].
\]
This work extracted is random because $\tilde U$ are random, but due to the particular way she chose the random angles, the work average is computable and given by
\[
\mean{W_N}=\tr[\H_N\R^{\otimes N}]-\tr[\H_NU\R_\cg^{\otimes N}U^\dag].
\]
Here, $\R_\cg=\sum_i p_i \pro{i}{i}$ is the coarse-grained state. Moreover, computing the average extracted work \emph{per copy} in the limit of large $N$ while optimizing over $U$  to maximize the extracted work yields an elegant formula called \emph{observational ergotropy},
\[
W_\C^\infty\equiv \lim_{N\rightarrow \infty}\max_U\frac{\mean{W_N}}{N}=\tr[\H\R]-\tr[\H\R_\beta].
\]
Here, $\R_\beta=e^{-\beta \H}/Z$ is a thermal state with temperature $\beta$ implicitly defined by requiring that its von Neumann entropy equals observational entropy~\cite{safranek2019b,safranek2019a,safranek2021brief,safranex2021generalized,strasberg2020first,riera2020finite,safranek2020classical,deutsch2020probabilistic,faiez2020typical,nation2020snapshots,strasberg2021clausius,hamazaki2022speed,zhou2022renyi,buscemi2023observational,modak2022observational,sreeram2023witnessing,schindler2023continuity}
\[
\Svn(\R_\beta)=S_\C,
\]
where $S_\C=-\sum_i p_i\ln p_i$ and we denoted the measurement basis as $\C=\{\ket{i}\}$~\footnote{The letter $\C$ for the measurement basis comes from ``coarse-graining'', which becomes more natural when we consider projective measurements (PVMs), given by a set of orthogonal projectors $\C=\{\P_i\}$, satisfying $\sum_i\P_i=\I$. Observational entropy is then defined as $S_\C=-\sum_i p_i\ln \frac{p_i}{V_i}$, where the volume of the corresponding ``macrostate'' is defined as $V_i=\tr[\P_i]$~\cite{safranek2019b}.}. The initial energy $\tr[\H\R]$ is unknown. Still, it can be well-estimated when the measurements performed on the system are local energy measurements~\cite{safranek2023work} and more loosely in the case of completely general measurements~\cite{safranek2023expectation}.

Observational ergotropy measures the amount of work Elizabeth will extract per copy on average by applying the unitary extraction~\eqref{eq:elizabeth_unitary_extraction}. Since it is inversely related to observational entropy, it is in her best interest to receive a measurement $\C$ together with probability distribution $\{p_i\}$ from Charlie that minimizes this entropy.

Charlie wants to help her. Thus, considering that he can choose from a set $\M$ of measurements that he can perform, he can try to find the best such measurement. In other words, he searches for a measurement that produces distribution $\{p_i\}$ that minimizes the observational entropy, solving
\[
\min_{\C\in\M} S_\C.
\]

Consequently, we update his goal as a certifier to find the best measurement and then send the source with a note of this optimal measurement basis and the corresponding experimentally measured probability distribution of outcomes to Elizabeth. This will guarantee that Elizabeth will extract the maximal amount of work given Charlie's abilities and her strategy.

Note that if Charlie is allowed to do any measurement, the minimum observational entropy he can attain is equal to the von Neumann entropy, which is reached when measuring in the eigenbasis of the density matrix~\cite{safranek2019b,buscemi2023observational}. Thus, von Neumann entropy sets the ultimate limit on the unitarily extracted work when allowed to perform general measurements in the protocol delineated above.

\sect{Review of the relation between observational and entanglement entropy} Interestingly, one can also find that entanglement entropy can be computed as the observational entropy minimized over local coarse-grainings~\cite{schindler2020correlation}
\[\label{eq:ent_entropy}
S_{\mathrm{ent}}=\min_{\C_A\otimes \C_B} S_{\C_A\otimes \C_B},
\]
where $\C_A\otimes \C_B=\{\ket{i}\otimes \ket{j}\}$ is the local measurement basis, and $S_{\C_A\otimes \C_B}=-\sum_{i,j}p_{ij}\ln p_{ij}$, where $p_{ij}=\abs{\braket{i,j}{\psi_{AB}}}^2$ is the joint probability of obtaining outcomes $i$ and $j$.

The minimum is attained for the local Schmidt basis $\{\ket{\psi_i}\otimes\ket{\phi_j}\}$, which is the product of eigenbases of the reduced states, $\R_A=\sum_i\lambda_i\pro{\psi_i}{\psi_i}$ and $\R_B=\sum_i\lambda_i\pro{\phi_i}{\phi_i}$.

\sect{Ergotropic interpretation of entanglement entropy} Combining the two results above gives rise to the ergotropic interpretation of entanglement entropy. 

Consider that Charlie can perform any measurement in a local basis (Fig.~\ref{Fig:ec} (b2)), i.e., the set allowed measurements is given by
\[
\M=\{\C_A\otimes \C_B\}.
\]
Additionally, consider that the state produced by the source is pure~\footnote{It is irrelevant whether Charlie or Elizabeth knows this --- the extracted energy will be the same either way.}. Suppose that Charlie can find the optimal measurement that minimizes the observational entropy (the local Schmidt basis measurement), then sends this information together with the corresponding optimal probability distribution $\{p_{ij}\}$ to Elizabeth. The maximum amount of work Elizabeth will extract per copy with her strategy (Fig.~\ref{Fig:ec} (c2)) is 
\[\label{eq:extracted_work}
W_\C^\infty=\tr[\H\R]-\tr[\H\R_\beta],
\]
where the temperature of the thermal state is defined implicitly by requiring that its von Neumann entropy equals entanglement entropy,
\[
\Svn(\R_\beta)=S_{\mathrm{ent}}.
\]
Thus, entanglement entropy determines the work that can be unitarily extracted from a source of pure quantum states characterized only by local measurements.

To see the second interesting interpretation of entanglement entropy, notice that $\R_\beta$ is the single-copy final state of the system, to which Elizabeth managed to cool it down by applying her unitary operations on it. To be more precise, while in a single realization, the entire protocol is unitary (see the extraction operation~\eqref{eq:elizabeth_unitary_extraction}), and the purity of the state is preserved, due to the lack of precise knowledge of the initial state, neither Charlie nor Elizabeth knows the final state. However, given that the entire procedure is random, upon many realizations of the work extraction protocol performed by Elizabeth, the final state is best described by the average state, which is the thermal state $\R_\beta$. What is meant by ``best described'' here is that there is no (arbitrarily complex) measurement that anybody can perform on states discarded by Elizabeth after extracting and distinguishing the outcomes from those produced by the state $\R_\beta$.

Thus, entanglement entropy also determines the minimum temperature to which a state characterized only by local measurements can be unitarily and predictably cooled down.

Finally, let us detail the word ``predictably'' used in the sentence above. Consider that Elizabeth could choose a different strategy that can, in principle, cool down the state even lower. Assume, for example, that she does not apply the random unitary $\tilde U$ first and instead applies the extraction unitary $U$ directly onto the state. If the state produced by the source is pure, and she is lucky to choose the unitary such that it transforms it to the ground state, the state's total energy is extracted. She, however, could not have known which unitary to apply to achieve this, given the limited knowledge she received from Charlie. The protocol explained above always works, meaning cooling down the system using this protocol leads to a thermal state with a predictable temperature with certainty while using only the information obtained from Charlie.

\sect{Non-pure states produced by the source}
The above interpretation held for global pure states produced by the source. Entanglement entropy is, however, defined only for global pure states. This is because, for mixed states, entanglement entropy can no longer be interpreted as a measure of entanglement. For example, a product state $\R=\R_A\otimes \R_B$, which clearly contains no entanglement, yields non-zero $\Svn(\R_A)$ and $\Svn(\R_B)$. There are other measures of entanglement applicable to mixed states, such as entanglement of formation~\cite{wooters1998entanglement}. 

However, a different quantity is relevant in the ergotropic context presented here. The interpretation of entanglement entropy rested on Eq.~\eqref{eq:ent_entropy}. This has a very natural generalization to mixed states. Minimizing the observational entropy over local measurement as follows,
\[\label{eq:ent_entropy2}
\min_{\C_A\otimes \C_B} S_{\C_A\otimes \C_B}\equiv \Svn+S_{AB}^{\mathrm{qc}},
\]
defines \emph{quantum correlation entropy} $S_{AB}^{\mathrm{qc}}$. Quantum correlation entropy is a measure of non-classical correlations similar to quantum discord~\cite{Bera_2018}, corresponding to entanglement entropy when the state is pure, $S_{AB}^{\mathrm{qc}}(\ket{\psi_{AB}})=S_{\mathrm{ent}}$. This quantity is identical to \emph{zero-way quantum deficit}~\cite{horodecki2005local}, and \emph{relative entropy of quantum discord}~\cite{modi2010unified,groisman2007quantumness,piani2011all}. The left-hand side given by the observational entropy, which is relevant for the optimal work extraction, is experimentally available given Charlie's abilities. Considering that only local measurements can be performed to characterize the source, but the states produced by the source are mixed, the maximal extracted work is again given by Eq.~\eqref{eq:extracted_work}. However, now the temperature of the thermal state is given implicitly by $\Svn(\R_\beta)=\min_{\C_A\otimes \C_B} S_{\C_A\otimes \C_B}.$

The two terms on the right-hand side of Eq.~\eqref{eq:ent_entropy2} are not experimentally available by themselves, considering they depend on the unknown state $\R$. However, they provide an interesting interpretation carried from the previous discussion: the sum of von Neumann and quantum correlation entropy determines both the maximal unitarily extractable work from a source characterized only by the local measurements, and the minimum temperature to which the states produced by an unknown source characterized only by local measurements can be unitarily and predictably cooled down.

\sect{Conclusion}
Entanglement entropy is used in many different contexts. In this paper, we exposed its elegant ergotropic interpretation. 
Entanglement entropy tells us how much energy we can unitarily extract from a source of unknown pure states when we have only local measurements to characterize it. 
Additionally, entanglement entropy determines how much we can cool down the states produced by this source with unitary operations.

\clearpage

\bibliography{main.bib}

\end{document}